\documentclass[orivec,copyright,creativecommons]{eptcs}
\providecommand{\event}{EXPRESS/SOS 2016}
\usepackage{breakurl}             % Not needed if you use pdflatex only.

\usepackage[nounderscore]{syntax}
\usepackage{amsmath}
\usepackage{amssymb}
\usepackage{array}
\usepackage{xypic}
\usepackage{tabularx}
\usepackage{pdfpages}\usepackage{breakurl}             % Not needed if you use pdflatex only.

\usepackage{graphicx}
\usepackage{times}
\usepackage{xspace}
\usepackage{subfigure}
\usepackage{ifthen}

\newcommand{\w}[1]{{\it  {#1}}}
\newcommand{\Alt}{~{\boldmath |}~}
  
\newcommand{\set}[1]{ \{ #1 \} }

\newcommand{\la}{\langle}
\newcommand{\ra}{\rangle}
\newcommand{\located}[1]{[\!| {#1} |\!]}

\newcommand{\ignore}[1]{}

\font\zmsy=zmsy10

\newcommand\boc{\mathop{\hbox to 11truept{\hfill\zmsy\char'060\hfill}}}

\newcommand\diaM{\mathop{\hbox{\zmsy\char'121}}}
\newcommand\boxM{\mathop{\hbox to 11truept{\hfill\zmsy\char'140\hfill}}}

\newcommand\opH{\boxM}
\newcommand\opP{\diaM}

\newcommand\fildia[1]{\opP}
\newcommand\myfilcir[1]{\mathop{\rlap{\hbox to 12pt
  {\hfill$\cir$\hfill}}\raise .55pt\hbox to 12pt
  {\hskip 0.4pt\hfill {$\scriptstyle#1$}\hfill}}}
\newcommand\filboc[1]{\opH}

 \newcommand{\cT}{{\cal T}}

\newtheorem{thm}{Theorem}
\newtheorem{lem}{Lemma}
\newtheorem{cor}{Corollary}
\newtheorem{obs}{ObservatDion}
\newtheorem{prop}{Proposition}

\newtheorem{defn}{Definition}

\newcommand{\proof}[1]{\par\noindent{\bf Proof:}~ #1 \hfill$\boxdot$}

\begin{document}

\title{Self-Similarity Breeds Resilience}
		\author{Sanjiva Prasad
			\institute{IIT Delhi }
			\email{sanjiva@cse.iitd.ac.in}
	\and
		Lenore D.~Zuck
		\thanks{%
			The second author was supported in part by NSF grant CNS-1228697}
		\institute{University of Illinois at Chicago}
		\email{lenore@cs.uic.edu}		
}

	\def\titlerunning{Self-Similarity Breeds Resilience}
	\def\authorrunning{S. Prasad \& L. D. Zuck}

\maketitle

\begin{abstract}
Self-similarity is the property of a system being similar to a part of itself.  
We posit that a special class of behaviourally self-similar systems exhibits a degree of resilience to adversarial behaviour.   
We formalise the notions of system, adversary and resilience in operational terms, based on transition systems and observations.
While the general problem of proving systems to be behaviourally self-similar is undecidable, 
we show, by casting them in the framework of well-structured transition systems, 
that there is an interesting class of systems for which the problem is decidable.   
We illustrate our prescriptive framework for resilience with some small examples, e.g.,
systems robust to failures in a fail-stop model, and those avoiding side-channel attacks. 
\vspace{1em}

\end{abstract}

%%% Local Variables: 
%%% mode: latex
%%% TeX-master: "main"
%%% End: 

% 15.Oct. SP:  Note to LDZ -- Blinded author info,  please  put page numbers

\section{Introduction}\label{INTRO}

% Motivation and main idea
Building systems that are resilient to actions of adversarial environments is an important software engineering concern. 
In this paper, we propose a class of systems whose resilience arises from a notion of {\em self-similarity}.  
An object is  said to be ``structurally self-similar'' if it is similar to a proper part of itself.
An important quality of the class of self-similar structures is that they are {\em scale-invariant}. 
% --- the properties of their part is the property of the whole (that is similar to the part).   
In analogy, we consider a class of systems that are {\em behaviourally self-similar} ---
 the behaviour of the system as a whole is ``equivalent'' to that of a part of the system ---
and develop a framework for showing how systems in this class are \textit{resilient to adversarial actions}.
The intuition behind our thesis is that if a part of a system is sufficient to exhibit the 
behaviour of the system as a whole, then the {\em rest} of the system
provides \textit{redundancy}, which in turn may provide resilience against a hostile environment.   
The notion of resilience is with respect to that of an {\em adversary}, a general concept 
pervading computing science, i.e., any way of choosing inputs or an environment  that 
can thwart a program from achieving its desired behaviour.

% Examples of self-similar systems
A trivial example of a behaviourally self-similar system is a constant signal.  
Its behaviour during \textit{any interval} of time is equivalent to its behaviour during an initial (arbitrarily small) interval, which is repeated {\it ad infinitum}.   
The signal can be considered to be resilient to an adversary that  determines {\em when} to sample the output value, 
in that it is able to map the adversary's sampling interval to a more convenient input (the initial interval) 
for which its behaviour has been defined.
Such ``delay-tolerance''  may also be seen in other time-independent functions.

% Recursive equations
The notion of behavioural self-similarity finds common currency in formal languages, in concurrency theory, as well as in programming. 
%
% Formal languages
In formal languages, we have a ready example of a construction that supports behavioural self-similarity, namely the Kleene star, as  $(e^*)^* \approx e^*$.  
Note that the construct $(~~)^*$ is semantically idempotent, a property that is often associated with fault-tolerance.  
%
% Concurrency
The replication operation in process calculi such as the $\pi$-calculus \cite{Milner-functions} is another example of self-similarity:
$!p \approx !p \| !p$, where $\approx$ denotes (behavioural/semantic) equivalence.    
It is also idempotent since $!!p \approx !p$, for any process $p$.
%
% Recursion
Indeed, in any recursive program scheme, we can find a syntactic part that behaves in a manner similar to the entire system.  
Consider, e.g., a recursive equation \( X \approx F[ X ],  \) for some non-trivial context $F[~]$.
By construction, $X$ is {\em a fortiori} operationally equivalent to the expression $F[X]$.   
The semantics attached to such recursive equations involves finding an appropriate fixed point,  usually the least fixed point,  preferably by a (finite) iterative process.  
Observe that placing any solution to this equation in the context $F[~]$ is a (behaviourally) idempotent operation\footnote{Finding the minimal fixed point context  helps us avoid ``needless redundancy''.}.  

% The main idea
From these examples, a na\"{i}ve idea arises linking system structure, behavioural self-similarity and resilience: 
If, assuming no adversarial action, a part $q$ of a system $p$ can behave as the whole system does, 
then this part can be considered to provide the core functionality of the system;
the rest of the system (the ``context'' $C[~]$, where $p \equiv C[q]$) serves to neutralise adversarial actions or transform the interactions of the adversarial environment $A$ with the system into a form which this ``core'' $q$ can digest, and thereby make the system behave as though adversarial action by $A$ were absent.   

% Notion of resilience.
\paragraph*{Resilience.} \ 
We say that a system $p$ is {\em resilient to an adversary} $A$ if its observable behaviour in the presence of adversarial action is 
equivalent to its behaviour in the absence of the adversary: 
\(  p \circ A \approx  p, \) where $\circ$ represents coupling the system $p$ with the adversary $A$\footnote{Note that $A$ need not be specified in the syntax of the language in which $p$ is expressed, and that the notion of coupling of the adversary to the system may be more general than the usual notion of parallel composition.}.
A somewhat similar formulation has been explored earlier by Liu, Joseph, Peled, Janowski and others \cite{LiuJ92,Peled-compositional,Janowski97}, but we believe our formulation is more natural (discussed in \S\ref{RELATED}). 
Now, if  $p$ can be expressed by  $C[q]$ as above, and in the absence of an adversary, $C[q] \approx q$,  we have, by transitivity, the desired resilience to the adversary $A$ arising from self-similarity.   
Resilience in this sense should not be equated with a notion of correctness;  
a system may be resilient even if it is not correct with respect to a given specification.
Note that if $\approx$ is a congruence, $C[C[q]] \approx C[q] \approx q$, showing the expected idempotence of $C[~]$ in countering adversary $A$.
% We however need to be more precise regarding what we consider equivalent behaviour, which requires a specification of what is {\em observable}.

\paragraph{Adversary model.} \ 
An adversary can be viewed as a way of forcing the program to face an unfavourable environment.  
The class of adversaries may be expressed in any of a variety of ways: as processes in a language, as automata or transition systems, as logical constraints on behaviour, etc.  
All that our framework requires is that the program coupled with the adversary is a transition system on which a reasonable notion of observational equality can be defined.  
We include in the class of adversaries a completely benign adversary, denoted $1_A$, who behaves as if there were no adversary present when coupled with any system.
If the adversarial model is specified as a transition system, we  require that it be {\em well-structured}, with $1_A \preceq A$ for any $A$ in the  class of adversaries.

We identify here some constraints on what an adversary can and cannot do. 
(i) Adversaries may act in ways completely unrestricted by the system.  
(ii) Adversarial moves, except for announcement of error,  are not directly observable.   
This is justifiable in that most adversaries are sneaky, not bruiting their actions, until and unless they wish to announce that they have defeated the system, i.e., { a denouement} \w{err}.  
(iii) An adversary cannot directly prevent the system from making any {\em observable} move by removing an enabled action. 
(iv) An adversary can, however, interact with the system, and make joint moves.  
These interactions too may not be directly observable, but may cause the system to (eventually) produce different observable effects from its normal behaviour.  

% Content and structure of the paper (put in the section references)
\paragraph*{Structure of the paper.} \ 
In the sequel, we develop this idea by formalising the notions of a system's behaviour, 
adversarial action, resilience and \textit{a methodology for proving a system to be resilient}, by factoring a system into its ``core'' and fault-tolerant context.
This is followed by a discussion of  related work (\S\ref{RELATED}).
Formalisation  (\S\ref{FRAMEWORK}) of both system and adversarial behaviour is done in the very general setting of {\em transition systems}. 
We employ a process calculus notation for expressing processes, and the associated structural operational semantics helps in relating structure to behavioural self-similarity.
In this paper, we use suitable notions  of {\em observation} and {\em equivalence}, namely, those of {\em barbs} and bisimulation \cite{Milner-barbed}, since we consider systems \textit{closed} when coupled with an adversary.
While other behavioural equivalences may be considered, we chose bisimulation 
since it is the finest extensional notion of equivalence of observable behaviour.  Proposition \ref{SOUND} expresses the soundness of our proposed methodology.

In general, proving bisimilarity of transition systems, and consequently showing a system to be behaviourally self-similar, is \textit{undecidable}.
However, we can show that this problem is decidable for an interesting class of systems 
(Proposition \ref{DECIDABLE}) by using the framework of\textit{ Well-Structured Transition Systems} (WSTS) \cite{Finkel-wsts}.
The WSTS conditions required for establishing decidability  seem to arise very \textit{naturally} from the constraints placed on the context and adversary.

We then illustrate our prescriptive framework for resilience with small examples (\S\ref{EXAMPLES}), such as 
systems robust to failures in a fail-stop model, and defeating side-channel attacks. 
In the examples in this paper, which progress from finite to finite-control to infinite state systems, 
we do not propose any new mechanisms for building resilient systems.  
We deploy the familiar armoury of devices --- redundancy, replication, retry,  and repetition --- for countering the arsenal at the adversary's disposal.
However,  our framework may be seen as providing a formal (methodological) justification of these constructions. 
While fault-resistance has earlier been shown using rigorous mathematical techniques, we believe that our use of the WSTS framework provides the basis for automated techniques for proving resilience, especially in the case of infinite-state systems.
We illustrate our approach by conveying only the intuition for the different examples, and omit  the tedious details of the  proofs.
In \S\ref{CONCLUSION}, we conclude with a discussion of our approach, its limitations, alternative frameworks for specifying and verifying resilience to adversarial behaviour, as well as some directions for future work.

\paragraph*{Methodology.}
Our proposed methodology is:
\begin{enumerate}
	\item We  identify a class of adversaries, with a {\em least} adversary $1_A$.
		Adversarial moves are not normally observable, except perhaps for a final barb $\w{err}$. 
	\item We decompose the system process $p$ into a core $q$ which provides the basic functionality
	and a (fault-resilient) context $C[~]$.  Thus $p \equiv C[q]$.
	In general, the context may be multi-holed.
	The context $C[~]$ should not alter the core functionality of $q$.  In particular, it should satisfy the following conditions:
	\begin{enumerate}
		\item $C[~]$ should permit $q$ to make any of its possible moves, i.e.,
		$q \longrightarrow q'$ implies $C[q] \longrightarrow C[q']$ and
		$q \Downarrow o$ implies $C[q] \Downarrow o$;
		\item If $C[~]$ and $q$ jointly make a move, then all of $q$'s possible barbs are preserved, i.e., 
		if $C[q] \Downarrow o$ and $C[q] \longrightarrow C'[q']$ then $C'[q'] \Downarrow o$;
		\item The context $C[~]$ (and its derivatives) should by itself contribute no observable barbs, i.e., $C[~] \not\Downarrow o$ for any $o$;
		\item No transition arising purely due to $C[~]$ disables the execution of the process $q$, i.e., if $C[~] \longrightarrow C'[~]$, then (i)  $q \Downarrow o$ implies $C'[q] \Downarrow o$ (since $C[q] \Downarrow o$), and (ii) 
		$q \longrightarrow q'$ implies $C'[q] \longrightarrow C'[q']$.
	\end{enumerate}
	\item We then specify the coupling $p \circ A$ of a process and an adversary as a transition system. 
	Formally, we will require that this transition system be a WSTS. 
	In particular, this composite transition system should exhibit the \textit{upward simulation property} (defined in \S\ref{FRAMEWORK}).
    \item To show resilience of $p$ with respect to the adversary $A$, we show that $q \circ 1_A \approx p \circ A$.
    This problem is decidable for WSTSs with \textit{certain additional properties}.
\end{enumerate}
We dub the conditions on the context and the adversary listed above the {\em self-similarity constraints}.

\paragraph{Beyond finite-state systems.}
While our examples in this paper are small,  our framework is not confined to dealing with finite (or finite-state) systems, for which it may be easy to show the required bisimilarity. 
Accordingly, we explore systems that provide sufficient structural properties to ensure that {\em bisimilarity is decidable}.     
We find that Well-Structured Transition Systems (WSTSs) \cite{Finkel-wsts} 
provide a framework in which we can formulate and \textit{verify} resilience by virtue of self-similarity.

Consider first a  \textit{simple version} of the framework:  
Structural inclusion of $q$ in $C[q]$ for a context satisfying the self-similarity conditions is an obvious candidate when defining an ordering relation ($q \|  r  \preceq C[q] \| r $).  A simple way to obtain the conditions on context $C[~]$ mentioned above is to constrain the hole(s) $[~]$ in context $C[~]$ to appear only at ``head'' or ``enabled'' positions.  
This allows $C[q]$ to simulate $q$, and if $C[~]$ has no observable actions, then every barb of $q$ is a barb of $C[q]$ and vice versa.   

Often the adversary itself can be formulated as a WSTS with a least element  $1_A$ representing the absence of an adversary.   
The composite transition system is obtained from those of the system and the adversary, and a \textit{pointwise combination} of the system and adversary orderings yields the desired ordering relation for the WSTS. 
We say that the composition with adversary $A$ is \textit{monotone} if $p \longrightarrow p'$ implies
$p \circ A \longrightarrow p' \circ A$, and $p \Downarrow o$ implies 
$p \circ A \Downarrow o$.  This is usually the case with parallel composition in most process calculi.

The self-similarity constraints on the context and adversary imply the following properties:
\begin{itemize}
	\item (Upward simulation) $q \longrightarrow q'$ implies $C[q] \longrightarrow C[q']$.  For monotone compositions with adversaries, this further implies
	$C[q] \circ A \longrightarrow C[q'] \circ A$.
	\item For an observable $o$,  $q \Downarrow o$ if and only if $C[q] \Downarrow o$.
	\item if $A \longrightarrow A'$ then for any $p$: $p \circ A \longrightarrow p \circ A'.$
\end{itemize}

What remains is to place  reasonable effectiveness constraints on the WSTSs  in order to ensure that bisimilarity is  decidable.   
We require that the states of the system and the class of adversaries  are {\em recursive sets} and that the ordering $\preceq$ is decidable.
We also assume that the transition systems are {\em finitely-branching}.  
In order to  ensure decidability, we require  that the transition systems satisfy a technical condition of having an {\em effective pred-basis},  and exhibit downward reflexive simulation (these definitions are recalled in \S\ref{FRAMEWORK}).

\subsection{Contributions and Related Work}\label{RELATED}
% Contributions 
We are unaware of any previous work where the notion of self-similarity has been wedded to that of
behavioural equivalence to formulate a notion of a system being resilient to actions by an adversarial environment.
Furthermore, we believe that the methodology we enunciate ---  construing a system as  being constructed of a core behaviour and a context for handling the actions of a formally defined adversarial environment --- is novel, as also casting them in the framework of Well-Structured Transition Systems \cite{Finkel-wsts}.
The structural decomposition of the system into core and fault-absorbing context seems natural and dovetails nicely with the WSTS conditions.  
As a consequence,  the bisimulation proofs become much easier.  Moreover,  the effectiveness conditions provide decidability, and thus in principle at least, support automated techniques for showing resilience that can work even for infinite-state systems. 
We also believe that our third example, which deals with a building block for converting communication over a non-FIFO channel with omission failures to a FIFO-channel with omission failures, has not been presented earlier  in its essential form.

The idea of formulating fault tolerance in terms of behavioural equivalence is not new \cite{LiuJ92,Peled-compositional,Janowski97}.    
The idea of a fault preorder, capturing the relative severity of faults, can be found in the works of Janowski, Krishnan and others \cite{Janowski97,Krishnan,LiuJ92}.  
Janowski, e.g., studies the problem of monotonicity of fault tolerance --- a system tolerant of faults higher in the preorder should tolerate faults lower in the preorder, but finds that this requirement does not square well with the standard notion of bisimilarity.
A similar observation is made by Krishnan \cite{Krishnan}, where he considers replicated systems  to model systems with synchronous majority voting.   Accordingly, a notion of bisimilarity parameterised by the fault model is proposed \cite{Janowski97,Krishnan}.  
In contrast, we believe that the notion of behaviour should be {\em uniform} and therefore formulate the notion of resilience to an adversary using a \textit{standard} notion of equivalence, e.g., weak (asynchronous) barbed bisimulation \cite{Milner-barbed,Amadio-asynch}.

Another major difference with these approaches \cite{Janowski94,Janowski97} is that they formulate the faulty 
versions of the system by incorporating the  anticipated faulty behaviour into definitions of the system.  
We see this is as unsatisfactory in that the adversarial behaviour has to be expressed concretely and within the syntax of the system (e.g., in the CCS formulation), thereby severely restricting the expressive power given to the adversary.  
It is also not a very satisfactory way of composing adversaries.  
Janowski uses the technique of incorporating the faulty version with the original system by providing a {\em redefinition of the original system} taking a non-deterministic choice of the two behaviours.
This  approach does not work well, e.g., with modelling a fail-stop model, because ``revenant'' processes become possible --- a system  which is supposed to have failed, rises Lazarus-like and exhibits some active behaviour.  
In contrast, our formulation uses a very general framework of transition systems, which may be specified and combined in any convenient manner.  
Indeed,  the syntactic formulation used for describing the example systems is only a convenient way for specifying a transition system and the constructive nature of fault-resilient transformations.   
 
There is also some similarity between our work and that of
Liu, Joseph, Peled  and others, in e.g., \cite{LiuJ92,Peled-compositional},  where they present frameworks  in which fault-tolerance is expressed by transforming a process with respect to the specification of a fault model:  $q$ is a $\psi$-tolerant implementation of $p$  if
$p \approx  T(q,\psi)$, for a transformation $T(\_, \_)$.   The juxtaposition of the recovery algorithm is viewed as a transformation that makes a process fault-tolerant. 
The connections between these logic-based ideas and our preliminary operational formulation deserves further study. 

\ignore{
\paragraph*{Structure of this paper.}   In \S\ref{FRAMEWORK}, we present the essential notions about transition systems,
barbed bisimilarity, and well-structured transition systems (WSTS), citing the particular decidability results that we use in our methodology.
We then provide some simple examples: resilience to side-channel attacks by introducing ``white noise'';
a server resilient to node failure (in a fail-stop model);  and transmission resilient to message reordering over a lossy channel.   These examples do not use any new mechanisms for resilience, though to our knowledge the third example has never been articulated earlier. 
In \S\ref{CONCLUSION}, we summarize the main ideas and indicate lines for future work.
}

\section{The Framework}\label{FRAMEWORK}
\newcommand{\Nat}{\mathbb{N}}
\newcommand{\nil}{\boldmath{0}}

\subsection{Transition Systems}

%\definition{
A \emph{transition system} $\cT= \la S, \longrightarrow \ra$ consists of a set of states $S$ 
and a transition relation $\longrightarrow \subseteq S \times S$.
A finite \emph{trace} $\sigma$ with respect to $\cT$ is a sequence of states $\la s_0, s_1, \ldots, s_n \ra$ such that  $s_i \longrightarrow s_{i+1}$ for all $0 \leq i < n$.   
An infinite trace with respect to $\cT$ can be considered to be a function  $\sigma: \Nat \rightarrow S$ such that for all $i \in \Nat $: $\sigma(i) \longrightarrow \sigma(i+1)$.  
We write $\sigma_1 \sigma_2$ to denote the concatenation  of traces, which in the case of $\sigma_1$ being an infinite trace, results in $\sigma_1$.
%}

The {\em successors} and {\em predecessors} of a state $s\in S$, are defined as $\w{Succ}(s) = \set{s' ~|~ s \longrightarrow s' }$ and  $\w{Pred}(s) = \set{s' ~|~ s' \longrightarrow s}$ respectively.  
The notations $\longrightarrow^n$, $\longrightarrow^=$, $\longrightarrow^+$ and $\longrightarrow^*$ are used for the $n$-step iteration, reflexive closure, transitive closure and reflexive-transitive closure of the transition relation $\longrightarrow$.  We use a similar notation for the $n$-step iterations, and closures of \w{Succ} and \w{Pred}.  
$\cT$ is {\em finitely-branching} if $\w{Succ}(s)$ is finite for each $s$.

Assuming a notion of {\em observable actions},  we define a {\em barb} as an observable action that a process has the potential to perform, written $p \downarrow o$ (where $o$ is observable)  \cite{Milner-barbed}.   
We will use ``weak barbs'' which depict the potential of a process to perform an observable action after making some ``silent'' transitions, i.e., $p \Downarrow o$ if for some $p'$,  $p \longrightarrow^* p'$ and $p' \downarrow o$.

A weak {\em barbed simulation} is a binary relation $R$ on processes such that if $(p_1, p_2) \in R$, then 
(1) if $p_1 \Downarrow o$ then $p_2 \Downarrow o$; and (2) whenever $p_1 \longrightarrow^* p'_1$ then
there exists $p'_2$ such that $p_2 \longrightarrow^* p'_2$ and $(p'_1, p'_2) \in R$.  
$R$ is a {\em weak	barbed bisimulation} if  both $R$ and its symmetric inverse $R^{-1}$ are weak barbed simulations.  
Processes  $p_1$ and $p_2$ are weakly barbed bisimilar, written $p_1 \approx_b p_2$, if they are related by some weak barbed bisimulation.  
Weak barbed bisimulation is not fine enough to distinguish processes that differ after the first communication (barb), when they interact with other processes, so the equivalence relation usually desired is a refinement that is preserved under parallel composition. 
\[  p_1 \approx p_2 ~\w{if} ~\forall ~q: ~(p_1 \| q) \approx_b (p_2 \| q) \]
However, for closed systems, it is reasonable to use weak barbed bisimulation as the notion of equivalence. 
%  if $\forall ~A: ~(p_1 \circ A) \approx_b (p_2 \circ A)

\paragraph{Process notation.}
In our examples, we employ a process calculus notation akin to a value-passing  CCS.  
Let $a$  represent a channel, $x$ a variable, and $v$ a value drawn from some set of values.
We assume without further specification that the language includes a syntactic category of expressions $e$, which contains in particular variables and integer-valued expressions.   
In our examples, expressions may also include tuples, and we assume a matching operation. 
Terms in the  language of communicating processes, typically $p, p_1, p_2$ are specified inductively by the following abstract  syntax:
\[ p::=  \nil  \Alt \overline{a}e.p \Alt ax.p_1 \Alt p_1 \| p_2 \Alt (a) p_1  \Alt p_1 + p_2 \Alt [e_1=e_2] p \Alt !p \Alt P(e_1, \ldots e_n)
\]
The process $\nil$ is inert.  
The prefix $\overline{a}e$ stands for the output of the value of $e$ on channel $a$, whereas  $ax.p$ stands for input of a value over channel $a$ with the value bound to $x$ in the continuation $p$.  
$ p_1 \| p_2 $ represents  parallel composition whereas $p_1 + p_2$ stands for non-deterministic choice between $p_1$ and $p_2$.  
The notation $(a) p_1$ describes the {\em restriction} operation on channels, i.e., channel $a$ is local to scope $p_1$.  Communication on restricted channels is {\em not} observable.  
$[e_1=e_2] p$ is a conditional matching operation.
$!p$ represents the replication of  process $p$, yielding as many copies of $p$ as desired, running in parallel. 
For convenience, we include parameterized (recursive) processes of the form $P(e_1, \ldots e_n)$.  

In a distributed system, we  associate processes with {\em locations}, written for instance as $\ell\located{p}$, where $\ell$ is a location constant. 
A context $C[~]$ is a process term with a hole $[~]$ in the place of a process term.  
We may also have multi-hole contexts.
We do not  present here the formal rules for the operational semantics of this language. 
Indeed, these constructs can be encoded in a core asynchronous calculus with replication and choice restricted to {\em guarded processes}.
We refer the reader to any standard presentation of such asynchronous calculi \cite{Honda-reduction,Amadio-asynch}
for the encodings and the operational semantics rules.

\subsection{Well-structured Transition Systems}

We now summarise some results about WSTSs \cite{Finkel-wsts}.

\noindent
A {\em quasi-order} or {\em pre-order}  $\la S, \preceq \ra$ consists of a set $S$ 
with any reflexive and transitive relation $\preceq \subseteq S \times S$.
$\la S, \preceq \ra$ is {\em well-ordered}  (henceforth a {\em wqo}) if for any \emph{infinite} sequence
$s_0, s_1, \ldots$ there  exist indices $i$ and $j$ with $i < j$ such that $s_i \preceq s_j$.  
Consequently, a wqo has no infinite strictly decreasing sequence, nor any infinite sequence of unrelated 
elements.  
It also follows that in a wqo, any infinite sequence $s_0, s_1, \ldots$  has an
\emph{infinite  non-decreasing subsequence}
$s_{i_0} \preceq s_{i_1}  \preceq s_{i_2} \ldots$ where $i_0 < i_1 < i_2  \ldots$.

\noindent
In quasi-order $\la S, \preceq \ra$, \emph{an upward-closed set} is any set $I  \subseteq S$ such that 
if $x \in I$ and $x \preceq y$, then $y \in I$.  Let $\uparrow x = \set{ y ~|~ x \preceq y}$, called the upward closed
set induced by $x$.  A {\em basis} for an upward closed set $I$ is a set $I_b \subseteq I$ such that
$I ~=~ \bigcup_{x \in I_b} (\uparrow x)$.

\noindent
\emph{Higman's Lemma} states that if $\la S, \preceq \ra$ is a wqo, then any upward-closed set $I$ has a {\em finite basis}.
The minimal elements of $I$ form a basis, and these must be finite, since otherwise, they would include an infinite sequence
of unrelated elements.   Further, any chain of upward-closed sets $I_0 \subseteq I_1 \subseteq I_2 \subseteq \ldots$ stabilises, i.e.,
for some $k$, $I_j=I_k$ for every $j \ge k$. 

\noindent
A \emph{Well-Structured Transition System} (WSTS)  $\cT = \la S, \longrightarrow, \preceq \ra$
consists of a transition system $ \la S, \longrightarrow \ra$ equipped with a well-ordered quasi-order
$\la S, \preceq \ra$ that satisfies \emph{weak upward simulation}:
For every $s$, $s'$, and $t \in S$, 
 if $s \longrightarrow s'$ and $s \preceq t$, then 
there exists $t' \in S$ such that $t \longrightarrow^* t'$ and $s'  \preceq t'$.

\noindent
A WSTS exhibits {\em downward reflexive simulation} if for  each $s$, if $s \longrightarrow s'$ and $t \preceq s$, then there exists $t'$ such that $t \longrightarrow^= t'$ and $t'  \preceq s'$,
i.e., either $t \equiv t'$ (0 steps) or $t \longrightarrow t'$ (1 step).

\noindent
A  WSTS $\cT = \la S, \longrightarrow, \preceq \ra$ has an \emph{effective pred-basis} \cite{Finkel-wsts} 
if there exists an algorithm that, given any state $s \in S$, computes
\w{pb}(s), a finite basis of $\uparrow \w{Pred}(\uparrow s)$,
i.e., minimal elements of the upward-closed set induced by the predecessors of states in the upward-closed set induced by $s$.
%}

\noindent
Backward reachability analysis involves computing $\w{Pred}^*(J)$ as the limit of a chain $J_i$, where $J_i \subseteq J_{i+1}$.  If $J$ is upward-closed, then this process converges, and $\w{Pred}^*(J)$ is upward-closed.
If a WSTS  $\cT= \la S, \longrightarrow, \preceq \ra$ has an \emph{effective pred-basis} and $\preceq$ is decidable,  then if any upward-closed $J$ is given via its finite basis, one can compute a finite basis of $\w{Pred}^*(J)$.

\noindent
The {\em covering problem} is, given states $s$ and $t$, to decide whether there exists a $t'$ such that $s \longrightarrow^* t'$ and $t \preceq t'$. 
The covering problem is decidable for  $\cT = \la S, \longrightarrow, \preceq \ra$ with an \emph{effective pred-basis} and  decidable $\preceq$.
% The proof involves computing a finite basis for $\w{Pred}^*(\uparrow t)$, and checking if $s$ is in it, which is possible since the  basis is finite and $\preceq$ is decidable. 

\noindent
If $\cT = \la S, \longrightarrow, \preceq \ra$  exhibits  downward reflexive simulation, and if $\w{Succ}$ is computable and $\preceq$ decidable, then one can compute for any $s$ a finite basis of $\uparrow \w{Succ}^*(s)$.  

\noindent
The {\em sub-covering} problem is to decide, given $s$ and $t$, whether there exists $t'$ such that $s \longrightarrow^* t'$ and $t' \preceq t$.
{\em Subcovering} is decidable for any WSTS which shows downward reflexive simulation,  if $\w{Succ}$ is computable and $\preceq$ decidable.

Putting the pieces together, we get the following method for proving whether a putatively fault-tolerant process  $p \equiv C[q]$ is
indeed resilient (or not) to an adversary $A$, where $C[~]$ denotes the fault-digesting context and $q$ its core functionality:
\begin{prop}\label{SOUND}
	A process $p$ is resilient to adversary $A$ while providing behaviour $q$ if $p$ can be expressed as $C[q]$ such that $C[~]$ and $A$ satisfy the self-similarity constraints, and $C[q] \circ A \not\Downarrow \w{err}$.
\end{prop} 
\proof{
Consider $C[q] \circ A$.  We wish to show $q \circ 1_A \approx C[q] \circ A$.
\begin{itemize}
\item {\em Upward Simulation:}  The adversary does not directly restrict any action of the system.
	Since $1_A$ represents the non-existence of an adversary, $q \circ 1_A \longrightarrow t$
	implies $t \equiv q' \circ 1_A$ for some $q'$.
	If $q \Downarrow o$, then clearly $q \circ 1_A \Downarrow o$ and
	also $C[q] \Downarrow o$.  If composition with the adversary is monotone, then 
	$C[q] \circ A \Downarrow o$.
	In general, this may not be the case (as will be seen in the second example of  \S\ref{EXAMPLES}).   If for a context $C[~]$, upward simulation is not satisfied, then that context \textit{fails to provide resilience} to the adversary $A$\footnote{The above-mentioned second example satisfies the WSTS condition only for certain contexts amongst contexts that satisfy the other self-similarity conditions 2(a)-2(d).}. 
\item {\em Downward reflexive simulation:}  
	The adversary or its interaction with the system does not cause anything that the core cannot do.
	Since the context $C[~]$ and the adversary do not contribute observables (except the ``denouement''), $C[q] \circ A \Downarrow o$  implies
	either $q \Downarrow o$  or that the process is {\em not} resilient to adversary $A$ (if $o$ is the barb $\w{err}$).  (The first example in \S\ref{EXAMPLES} involves an
	adversary that attempts to make observable the barb $\w{err}$.)
	Now consider the moves that $C[q] \circ A$ can make.  These may be 
	\begin{enumerate}
	\item a move due to $q$:  $C[q] \longrightarrow C[q']$.  This is downward simulated by $q \longrightarrow q'$ (and hence $q \circ 1_A \longrightarrow q' \circ 1_A$).
	\item a move due to only $C[~]$: $C[q] \longrightarrow C'[q]$.  This is downward simulated by $q$ making no move.  Moreover, $C[q] \Downarrow o$ iff  $C'[q] \Downarrow o$ iff $q \Downarrow o$.
	\item a move due to $A$:  $C[q] \circ A \longrightarrow  C[q] \circ A'$.  Since the adversary's moves are unobservable, 
	this is downward simulated by $q \circ 1_A$ making no move, since $1_A \preceq_2 A'$.
	\item a move involving both $C[~]$ and $q$: $C[q] \longrightarrow  C'[q']$: this is simulated by 
	$q \longrightarrow q'$, and $C'[q'] \Downarrow o$ iff $q' \Downarrow o$, since $q' \preceq C'[q']$ and $C'[~]$ does not contribute any observables and
	does not inhibit any observables of $q'$.
	\item A move involving the adversary {\em and} the fault-handling context $C[~]$:
	$C[q] \circ A \longrightarrow C'[q] \circ A'$.  Again, since this move is unobservable,
	this is downward simulated by $q \circ 1_A$ making no move, since $q \circ 1_A \preceq C'[q] \circ A'$  for $C'[~], A'$.
	\end{enumerate}
\end{itemize}
}

Thus, our prescriptive framework of crafting self-similar processes and formalising adversarial behaviour
operationally yields the desired resilience.   
If the two systems are not barbed bisimilar, then the process is not fault tolerant, and a counter-example may be found.
Another way of looking at the definition of resilience is that adversarial moves keep the composite system within the same behavioural equivalence class.

\paragraph{Dealing with infinite state systems.} \ \ 
Since the state spaces may potentially be infinite, there may not be an effective method to {\em prove} that 
$C[q] \circ A$ and $q \circ 1_A$ are bisimilar.   In order for this question to be decidable, we place the restrictions mentioned earlier, namely that
 the WSTS has an effective pred-basis, that the successor relation is effectively computable,
and that the conditions on adversarial moves ensure reflexive downward simulation. The ``self-similar subterm''  orderings are clearly decidable. 

\begin{prop}\label{DECIDABLE}
The problem of deciding whether a process is resilient to an adversary is decidable if in addition to  satisfying the self-similarity constraints, the coupled transition system has an effective pred-basis, and the successor relationship is effectively computable.
\end{prop} 
\proof{
Suppose $q \circ 1_A \longrightarrow t $.   We need to show that there is a $t'$ such that $C[q] \circ A \longrightarrow^* t'$ and $t  \preceq t'$.
This is an instance of the {\em covering problem}  for $C[q] \circ A$  and $t$, which is decidable under the assumptions.
{\em Proof technique:}
Compute $K_b$, the finite basis of $\w{Pred}^*(\uparrow t)$,
and check if $C[q] \circ A \in \uparrow K_b$. \\
Now suppose $C[q] \circ A \longrightarrow t$. We need to show that there is a $t'$ such that $q \longrightarrow^*t'$ and $t' \preceq t$.
This is an instance of the sub-covering problem for $q$ and $t$, which is decidable when context/adversarial/joint moves are simulated
downwards in 0 steps, and successors are effectively computable.  
{\em Proof technique:}
Compute $K_b$, the finite basis for $\w{Succ}^*(q \circ 1_A)$.  Check if $t \in \uparrow K_b$.
}

\section{Examples}\label{EXAMPLES}

In this section we put our proposed methodology to test by applying it in situations that demand resilience of a system to an adversary.  
Our examples proceed from  finite to finite-control to infinite state systems.
In each case, we apply the methodology of identifying  the core computation that would have sufficed in the absence of an adversary.
We then identify an adversary and show how by constructing an absorptive context satisfying the self-similarity conditions presented earlier,  one can defeat the adversary.
The wqo notion used is usually simple, though the later examples, involving resilience in distributed systems, motivate the need for more flexible notions 
than simple  embedding of a process into a context.   
However, they retain the essential semantic requirement that the context preserves the ability of a process to perform its actions, and  that the context neither contributes any observable actions directly, nor does it take away the observables of the core process.  
It can at best interact with other parts of the context and/or with the adversary.  

\subsection{White Noise to defeat Side-channel Attacks} 

Let $c$ be a deterministic finite computation which generates an observable result $m$ in $n$ steps:   $c \longrightarrow^n c' \downarrow m$.  
Normally, observers cannot see the number of steps taken by $c$. 
Let us now consider an adversary that can see in addition to this outcome, the ``side property'' of how many steps were taken to termination\footnote{Other side properties such as heat generated, or power consumed could also be monitored.}.  
If it observes that $c$ terminates within $n$ steps, it flags $\w{err}$.  
A class of step-counting adversaries can be coded as
$A(i,k)$ for $k \geq 0$, with the behaviour of the composite transition system $c \circ A(i,n)$ given by the following rules:
\[
\frac{p \longrightarrow p'}{p \circ A(i,n)  \longrightarrow p' \circ A(i+1,n)}
~~~
\frac{p \downarrow m}{p \circ A(i,n) \downarrow m}
~~~
\frac{p \downarrow m}{p \circ A(i,n) \downarrow \w{err}} ~~(i \leq n)
\]
Note that the adversary does not suppress any observable of $p$, and makes no observable moves except signalling $\w{err}$.
In the absence of an adversary monitoring the side-channel, we have  $c \circ 1_A \Downarrow m$, but
in the presence of such an adversary, $c \circ A(0,n) \Downarrow \w{err}$ as well.
Thus if $c_1$ is a program behaviourally equivalent  to $c$  but which takes more than $n$ steps to terminate, this adversary can distinguish between $c$ and $c_1$ as
$c_1 \circ A(0,n)  \not\Downarrow \w{err}.$

We now justify the correctness of the method of interleaving a computation with an
indeterminate number of NOPs to defeat such side-channel attacks. 
Let \w{WhiteNoise} be a computation that consumes cycles, but does {\em not} generate any  observables, e.g., $\w{WhiteNoise} \longrightarrow \w{WhiteNoise}$.
\w{WhiteNoise} does not  suppress or alter any normal observables of computations running in parallel/interleaved with it.
Consider the context \(  C[~] = \w{WhiteNoise} \|_f [ ~], \)  and suppose $\|_f$ is a weakly
fair (nondeterministic\footnote{If the interleaving performs a deterministic number of \w{Whitenoise} steps, then by a small extension to the adversary class, one can mount a side-channel attack.}) interleaved implementation of parallel composition that executes at least one step of \w{WhiteNoise}.
Now note that while  \( C[ c ] \circ  A(0,n) \Downarrow m \), it is no longer the case that \( C[ c ] \circ  A(0,n) \Downarrow \w{err}. \)
Thus $A(0,n)$ cannot distinguish between $C[c]$ and $C[c_1]$.  

It is straightforward to show that $C[~]$ and $A(0,n)$ satisfy the self-similarity conditions.  
We cast the composite system as a WSTS, using the ``self-similar subterm'' ordering on processes.
In particular,  we  consider as minimal elements (which form a finite basis) all processes $c'$ such that $c \longrightarrow^* c'$.  
The \textit{effective pred-basis} is then easy to compute.
It is therefore easy to prove that $C[ c ] \circ A(0,n) \approx c$.
Note also that this context $C[~]$  is idempotent; reiterating it, as in $C [ C [ c ] ]$, does not provide further security against this side-channel attack.

\subsection{Replicated server}

The next example involves finite-control, and justifies the use of repetition to address multiplicity of requests and spatial replication to counter failure. 
Consider a one-time provider of a file $v$: ~~
\( \w{OTP} =  \overline{a}v,  \)
where $a$ is the channel on which $v$ is sent.
Similarly, a basic client  is rendered as 
\( \w{BC}_i = ax. \overline{d_i}x, \)
which receives some file $x$ on the channel $a$, and delivers it
to the client's application layer, written as $\overline{d_i}x$. 
The single-request client-provider system is written as
\( \w{Sys}_1 = (a) ( \w{BC}_1 \|  \w{OTP} ). \)
The only observable barb of $\w{Sys}_1$ is the unrestricted send action $\overline{d_1}v$.

If the file provider has to deal with more than one client, 
or if the client repeatedly requests the file,  we need an ever-obliging
{\em responsive} server,
represented as a process that can {\em repeatedly}  send file $v$ on the channel $a$. 
\[ \w{Server} = \ell_1\located{  !~[  \overline{a}v ] }.  \]
% where $!~[ ~ ]$ can be coded as   
% \[ C[ ~ ] =  (b) !(bx. [ ~ ] \| \overline{b}x) \| \overline{b}t \] for  some $t$.
Typically, the server is located at some site $\ell_1$, written in a distributed calculus (e.g., \cite{Riely-distributed}) as $\ell_1\located{\ldots}$, which may be different from the client's site.    
 For simplicity we have the \w{Server} located at $\ell_1$  repeatedly sending the file over $a$ 
 to whoever wishes to receive it\footnote{Typically this is coded (in a $\pi$-calculus) as a client sending a request to a server, sending a private channel over which it wishes to receive the file.}.
In the distributed setting, note that  $p \longrightarrow p'$ implies $\ell_1\located{p} \longrightarrow \ell_1\located{p'}$,  and $p \downarrow o$ implies $\ell_1\located{p} \downarrow o$ only when site $\ell_1$ is ``up'' (locations do not figure in the observable barbs).
%\w{Server} is a replicated process that is triggered by an input on local channel $b$; the i%nitial trigger is $ \overline{b}t$, which is {\em recycled} in every replication.  
Observe that  the context in which \w{OTP} is placed  contributes no barbs, and has a hole in a position that is enabled.   Consider the new system:
\[ \w{Sys}_2 = (a)( \w{BC}_1 \| \ldots \w{BC}_n \| \w{Server} ) \]
% where $\w{BC}_i = ax. \overline{d_i}x$.
The observable barbs are $\overline{d_i}v$, for $i \in \{1,\ldots,n\}$.
Since $a$ is restricted, the send actions on $a$ do not contribute any observable barb.  
This construction also handles the case when the clients repeatedly request a value, i.e., when 
$\w{BC}_i ~=~ ! (ax. \overline{d_i}x)$.

Now consider  an adversary $A$ that can cause location $\ell_1$ to fail, taking the server down permanently in the fail-stop model (see e.g., \cite{Amadio-localities,Riely-distributed})
after which no client can receive any file from the  \w{Server}.  Observe that the requirement that $\ell_1$ needs to be ``up'' for a barb to be observable means that the coupling with the adversary is \textit{not} monotone.  The adversary can be modelled as a transition system with states that represent the set of locations that are ``up'' and the transition $\{\ell_1\} \longrightarrow \{\}$ to model the failure of $\ell_1$.
It is now  possible to have trace suffixes in which
 \( \w{Sys}_2 \circ A ~\not\Downarrow \overline{d_i}v  \) whereas \( \w{Sys}_2 \circ 1_A ~\Downarrow \overline{d_i}v, \) for some value(s) of $i$.
Thus, $\w{Sys}_2$ is not resilient to an adversary that can cause a single location to fail.

We build a server resilient to a {\em single node failure} by {\em replicating} the responsive file provision on two sites, $\ell_1$ and $\ell_2$.
Fault tolerance is provided by using the two hole context
\[ C_{rep}[~] =  \ell_1\located{ ! [ ~ ]_1 } ~\|~ \ell_2\located{ ! [ ~ ]_2 }  \] that places a
process  in two holes both at head positions, and satisfying all the requisite self-similarity conditions for the context, yielding a {\em replicated responsive server}:
\[ \w{RepServer} =  C_{rep}[ \w{OTP} ] ~\equiv~ \ell_1\located{! \overline{a}v}  \| 
\ell_2\located{! \overline{a}v}. \]  
The server process located at $\ell_i$ can execute only if that location has not failed (fail-stop model of failure) --- it provides the value $v$ on channel $a$ while its site is up.  Let
\[ \w{Sys}_3 = (a)( \w{BC}_1 \| \ldots \w{BC}_n \| \w{RepServer})
\]
Client $\w{BC}_i$ can read a value on channel $a$ either from the server at $\ell_1$ or $\ell_2$, whichever is up; if both are up, it obtains the value from either one (the location of the server is not observable).
So if either or both $\ell_i$ are up, then $\ldots \w{BC}_i \ldots \| \w{RepServer} \Downarrow \overline{d_i}v$.

The adversary $A$, which is able to cause {\em at most one node} to fail, can be modelled by a finite state machine, which {\em may} make a transition from a state in which both  $\set{ \ell_1, \ell_2} $ are up, to states where $\ell_1$ (respectively $\ell_2$) is down, and then  remains in that state  (modelling fail-stop of at most one of the two sites).    The benign adversary $1_A$ can be modelled as a single-state FSM (with a self-loop).  
It is easy to formulate the composite system as a WSTS, by using the self-similar subterm ordering on processes, and the obvious trivial ordering on the adversary FSMs.  
It is also easy to exhibit the minimal elements and the effectiveness conditions for this system.   Thus we can prove that $\w{Sys}_3 \circ A \approx \w{Sys}_2 \circ 1_A$.  
This being a finite-control system, the proof is quite easy.  Our   WSTS-based framework is also able to handle extensions to the system to permit persistent clients, which repeatedly request the file and deliver it {\em ad nauseum}. 

Replicating \w{RepServer} again by placing it in $C_{rep}[~]$ serves no purpose with respect to  an adversary that can cause only one of the $\ell_i$ to fail\footnote{This is true  even when the language permits nested sublocations.}.  However, if the adversary can cause  $k > 1$,  to fail,  then a replication context should place the server at least $k+1$ locations.

\subsection{Reliable transmission}

We now address resilience to adversaries that make communication channels unreliable.  
Instead of presenting yet another verification of the ABP protocol \cite{ABP1969}
and its bisimulation proofs, which have been published several times before (e.g., \cite{LarsenM92}),
we  describe a small protocol for communication between a client 
$R$ and a server $\w{Sr}$ over  a channel $c$ that may arbitrarily reorder messages and omit messages. 
This protocol can be used as a basic building block within a larger protocol that builds a 
FIFO channel over a non-FIFO layer \cite{AfekAFFLMWZ94}.  
To our knowledge, this construction  has not been earlier presented in as simple a formulation.
 Its core is similar to the ``probe" construct of Afek and Gafni \cite{GafniA88}.
 % Please check this is the correct reference...
 
 The server is extremely simple: it receives a request on  channel $b$, and sends a message on channel $c$. At any point of time, it may segue to sending another value $v'$.
 \[ \w{Sr}(v) = (b. \overline{c}v. \w{Sr}(v)) + \w{Sr}(v'),  \]
 where $v' \neq v$ and $v,v' \in D$, some set of values (which we assume for convenience is of cardinality $k$).
 The server is representable as a (parameterised) non-deterministic finite-control machine.
 
 A  simple client $R_s$ can be expressed as:
 \( R_s = \overline{b}. cx.\overline{d}x.R_s, \) where
 $R_s$  sends requests on $b$ and, on receiving a value on channel $c$, delivers it  over channel $d$ and repeats.
 If channel $c$ does not omit or reorder values, $R_s$ will produce all the values sent by the server in FIFO order.  
  If, however, $b,c$ are {\em lossy} channels, then some requests (or responses) may be lost in transmission.  $R_s$ may therefore get stuck waiting for messages. 
  We make the client more persistent, modifying $R_s$ to $R_o$, 
 \[ R_o = ! [ \overline{b} ]_1 ~\|~ [ cx]_2 . [\overline{d}x ]_3 . R_o \]
 which decouples the (repeated) request sending on $b$ from the receipt of messages on $c$. 
 Since it works with lossy channels, $R_o$ may omit delivering some messages,  but all delivered messages appear in the order in which they were sent.  Note that we have placed the three communication actions in three distinct ``holes'' ($[~]_1, [~]_2,[~]_3$ ).

If, however, the channel $c$ can reorder messages, it is possible to confuse messages corresponding to earlier requests with those for later requests.  While this problem can be addressed by placing serial numbers on the messages, an interesting question is whether it can be solved without that mechanism.  Accordingly, we modify the client:
\[
\begin{array}{rcll}
R_{ro}(p, n_1, \ldots, n_k) & = & [ \overline{b} ]_1 .R_{ro}(p+1, n_1, \ldots, n_k) \\
                                        &     & + R'(p, n_1, \ldots, n_k)   \\ \\
R'(p, n_1, \ldots, n_k)  & = & [cx ]_2.  \textbf{case}~x~\textbf{of} & \\
                                   &    &        \vdots  & \\
                                   &      &    ~v_i: ~ & \textbf{if}~(n_i > 0)~\textbf{then}
                                    \\
                                   &   &                    &  R_{ro}(p-1, n_1, \ldots, n_i-1, \ldots, n_k) \\
                                   &    &                   & \textbf{else}~[ \overline{d}x ]_3 . R_{ro}(0, p-1, \ldots, p-1) \\
                                   &    &    \vdots \\
\end{array}
\]

The client initiates a {\em round} of the protocol by sending a request to the server on channel $b$. 
Since the responses from the server may be reordered or dropped by the response channel $c$, the client has no way of knowing whether a response it receives is acknowledging its current request (a ``fresh" message), or whether it is a response to a previous request.  However, by pigeon-holing, if it receives more responses than pending unanswered requests from the past, it knows that the server has acknowledged its most recent request.  
For this it keeps a variable $p$, which is the number of requests sent in this round, and
bounds the number of responses it requires for each value so that when it receives a response on $c$, it can determine that it has got enough responses to safely conclude that the value is a fresh one.   Note that in our soution, \textit{the server remains unchanged}, and in the client, all the parameters take \textit{non-negative integral} values.

Let $A_{ro}$ be an adversary that may reorder and omit messages (the interesting ability of the adversary lies in its being able to reorder responses on channel $c$), and let $A_o$ be an adversary that may only omit messages on channel $c$.   The equivalence to be established is:
\[  (b)(c) ( \w{Sr} \| R_{ro}(0, 0, \ldots, 0) ) \circ A_{ro}  \approx  (b)(c) (\w{Sr} \| R_o) \circ A_{o} \]

One may notice that the process $R_o$ is not exactly syntactically embedded within a context in $R_{ro}$.  However, the essence of the self-similarity constraints is met as every communication action comes into an enabled position at exactly corresponding points marked by the holes $[~]_1, [~]_2,[~]_3$.  
Thus, the resulting ordering on processes $\preceq$ would (while still being decidable) be more complex, but nonetheless adheres to the self-similarity requirements and those of being a WSTS.  
In defining the states of the coupled transition system we consider the states of the
server $\w{Sr}$, and that
of the client  $R_o$ and $R_{ro}$.    In identifying the latter,  we demarcate as significant the three marked holes  as points where to pin  control.  The start of a ``round'' is a significant point where the control of the $R_o$ process is matched with that of the $R_{ro}$ process.  
We take into account the parameters of the  $R_{ro}$ process in framing the $\preceq$ relation.  Finally, we consider the channel states, adapting the subword ordering that  has been used for lossy channels in order to deal with lossy-reordering channels.   The details are omitted here, but WSTS technique provides a novel way of proving {\em operationally} the correctness of a protocol that has an unbounded state space. 

\ignore{
%  Write more here???  CHECK THIS
Consider now that full blown problem --- that of $R$ trying to learn a sequence of data items that $S$ has, and
writing them one by one. We briefly outline how the simple process will be used to accomplish that. 
Recall that process  $(S \| R_o)$  does very little -- it allows $R$ to initiate a phase where
$R$ requests a ``fresh" message from $S$ until it receives and writes this message.   It is easy to construct
now a system where the requests are parameterised with a parity bit, just like in the Alternating Bit Protocol, 
so that requests belonging to one phase can be differentiated from requests of the previous phase or the next 
phase. 

This, however, doesn't indicate when $S$ sends the next element in its sequence. The main problem is that $S$ cannot tell new requests from old ones. Note that a parity bit on the requests helps
not -- suppose that $S$ has received 20 requests for items in even positions, and had sent 20 such responses,
the last one of the $i^{th}$ item. How many additional such requests should $S$ receive before it knows that
$R$ is trying to learn the $(i+2)^{nd}$ item?  

But here we can use our construct in the reverse direction --- $S$ can run the reverse protocol, asking $R$ to acknowledge
it had learnt the $i^{th}$ data item, and when it receives enough confirmation (more than the number of pending  requests for
acknowledgement when $S$ initiated the current round)
it can be sure that $R$ is ready for the $(i+2)^{nd}$ item.
}

By orienting this protocol in both directions, one can, at the price of counters (for the pending and new messages)  obtain an implementation of a FIFO channel over a lossy reordering communication channel. 
The only  way we know to avoid the counters is using sequence numbers on data items, as in \cite{Stenning},  but this implies an infinite message alphabet. 
 For practical purposes, one usually assumes that sequence numbers can be recycled (in a ``sliding window" fashion) under the belief that the channel does not deliver an ``ancient" message.   
 On similar lines, if one assumes  a bound on the number of messages the channel may delay, then our counters can be limited by that bound.  
 This may be a reasonable assumption since, as shown in \cite{AfekAFFLMWZ94},  in any   implementation of a FIFO channel over a  lossy reordering one with a finite message alphabet, the more the messages that the channel delays, the more the messages that need be transmitted.

\section{Conclusion}\label{CONCLUSION}

The question of whether a specified program behaviour can be achieved versus an arbitrary adversary is, of course, undecidable in general, with several famous impossibility results.  Even the question of whether a given program is resilient to some particular adversary is in general not decidable. 

What we have sought to do is to formalise an intuitive connection between resilience to an adversary and the self-similarity in the structure and behaviour of a program.
It is a prescriptive framework, and we recognise that there are several other ways of correctly constructing fault-tolerant systems that do not fit this methodology. 
The program is constructed in terms of its core functionality (that generates the observables) and an absorptive context that soaks up the adversarial actions, but otherwise neither contributes nor detracts from the observable behaviour of a program\footnote{This factoring of a program into core and context cannot in general be automated.}.
While in our framework, we require that the adversary's moves are not directly observable (except in a denouement),  the interaction between the adversary and the program may  result in different observables from the normal execution. 
However, a fault-resilient program exhibits no difference.

Another way of thinking about the framework we have proposed is in terms of {\em abstractions} that operate as follows:
consider the traces of the system by itself, and also those of the system composed with the adversary.  
Consider the  equivalence relation that arises from an abstraction function which has the property of ``stuttering'' over moves by the adversary.  
A system may be considered tolerant of an adversary if every adversarial transition is within an induced equivalence class. 
While there is a certain simplicity to such an account, our proposed framework is richer in at least some respects:
First, it is able to account for moves made in conjunction between the adversary and the system (the interactive moves).  
Second, it is able to relate the structure of the system with its behavioural self-similarity, and capture the intuition that the seemingly redundant parts of a program provide resilience against adversaries.  

We are unaware of any published work on relating self-similarity with resilience.
While there has been a body of work in which adversaries have been classified according to a partial order based on the severity of their disruptive capabilities, and relativising the notion of  behavioural equivalence based on the adversary model, we believe this is less elegant or compelling than an account where a \textit{standard} notion of behavioural equivalence is used.
While we have used barbed bisimulation, other notions of equivalence may be appropriate in certain settings, and may provide easier methods for proving resilience. 
Another shortcoming of the previous approaches is that the adversarial model was incorporated into the syntax of the processes, whereas our approach supports more eclectic styles of specifying the transition system of processes in the context of an adversary. 

Further, while the previous approaches seem to be confined to finite (or at best finite state processes) -- where it is possible to find appropriate bisimulation relations, we are able to deal with a class of infinite-state systems by couching the problem in a WSTS framework. 
Although there has been substantial work on WSTSs,  our use in proving systems to be fault-resilient seems to be novel.  
We believe that one can in this framework also deal with composite adversarial models by combining the fault-resilient contexts  in novel ways (e.g., embedding one kind of resilient context inside another), although this is beyond the scope of the current paper. 

As we have presented very simple examples and introduced no new mechanisms for resilience, it may seem that the results are unsurprising.   However, it is satisfying to not only confirm the correctness of intuitive constructions and folklore but also find an effective basis for demonstrating their correctness.  It would be interesting to study whether our framework also provides a way of discovering \textit{minimal} constructions for resilience against particular adversaries. 

Of course, our proposed methodology  should be put to further tests.  For instance, it would be interesting to see whether techniques for building resilience in storage systems such as RAID \cite{RAID} would be amenable to these techniques.
A major area which needs addressing concerns cryptographic protocols and resilience in the face of cryptanalytic adversaries.  The challenge there is to find the right notion of structural orderings, and recognising the contexts that provide resilience.  
Another domain that deserves greater study is coding theory and the use of redundant bits to provide error-correction.
It is not obvious in such techniques whether  convergence (reaching fixed-points) can be obtained. 

In the future, we would also like to explore conditions under which, given a specification of the operational behaviour of the adversary, and the core functionality, it may be possible (if at all) to {\em synthesise} the fault-resilient context.   
We would like to develop a framework analogous to those of Liu, Joseph et al. where fault-tolerant versions of systems are developed using a stepwise refinement methodology, and transformations dealing with combined fault models.   

In the current paper, we have used an operational framework; 
in the future, we would like to explore a logic-based formulation, using e.g., knowledge-based analysis techniques to reason about resilience in the same way that correctness of distributed systems/protocols has been studied \cite{HalpernZ-little}. \\

\noindent
\textbf{Acknowledgements.}  We would like to thank the anonymous referees for their helpful suggestions in improving the paper.  In particular, we would like to thank them for being generous in their assessment while being eagle-eyed in spotting mistakes in the submission.

\bibliographystyle{eptcs}
\bibliography{SelfSim}

\end{document}